# Removal of Pollutants by Atmospheric Non Thermal Plasmas

*Ahmed Khacef[1*], Jean Marie Cormier[1], Jean Michel Pouvesle[1], Tiep Le Van[2]*

[1]GREMI, CNRS-Université d'Orléans, 14 rue d'Issoudun, B.P. 6744, 45067 Orléans Cedex 2, France.
 e-mail : ahmed.khacef@univ-orleans.fr
[2]Laboratory of Catalysis, IAMS Vietnam Academy of science and Technology, 1 Mac Dinh Chi Str. Ho Chi Minh City, Vietnam

## Introduction

The international regulations for exhaust gases from cars and from industry restrict the contamination to values difficult to handle with conventional removal technologies like thermal and catalytic oxidation or adsorption. Most of them are in the range of few hundreds of ppmv and below. At these low contamination levels a catalytic oxidation requires a substantial supply of thermal energy to be effective. As an alternative, the application of non-thermal plasma (NTP) instead of thermal energy for removal of toxic components has been demonstrated and received growing attention. After initial work on military toxic wastes, an increasing number of investigations have been devoted to the treatment of nitrogen oxides (NOx) and sulphur oxides (SOx) in flue gases, and to the decomposition of volatile organic compounds (VOCs) [1-8]. Typical examples are hydrocarbons, chlorocarbons and chlorofluorocarbons (CFCs). Contamination of exhaust air streams with gaseous hydrocarbons or organic solvent vapours occurs in many industrial processes, e. g. in chemical processing, in print and paint shops, in semiconductor processing as well as in soil remediation and water treatment.

In the present paper, we present results on the application of non thermal plasmas in two environmentally important fields: oxidative removal of VOC and NOx in excess of oxygen. The synergetic application of a plasma-catalytic treatment of NOx in excess of oxygen is also described.

## Non thermal plasmas

Plasma is a partially or fully ionized gas consisting of electrons, ions, atoms, and molecules. For the quantitative description of the plasma, the term of temperature is usually used. Thermal plasma (TM) is in a state where almost all its components are at thermal equilibrium. In NTPs, temperature (i.e. kinetic energy) is not in thermal equilibrium, and differs substantially between the electrons and the other particles (ions, atoms, and molecules). In this sense, an NTP is also referred to as "non-equilibrium plasma" or "cold plasma". NTPs may be produced by a variety of electrical discharges (pulsed corona discharge, barrier discharge, glow discharge) [9-12] or electron beams irradiation [1, 13]. The electron beam technique which is very efficient for removal of pollutants such NOx and VOCs has been first used. However, pulsed discharges (corona and dielectric barrier discharge-DBD) are much more suited than e-beams for some industrial and domestic applications because of their high selectivity, moderate operating conditions (atmospheric pressure and room temperature), and relatively low maintenance requirements resulting in relatively low energy costs of the pollutant treatment.

The basic feature of NTP technologies is that they produce plasmas in which the majority of the electrical energy primarily goes into the production of energetic electrons instead of heating the entire gas stream. These electrons can activate the gas molecules by collision processes and subsequently initiate a number of reaction paths generating additional O, OH or $HO_2$ radicals for decomposing pollutants. However, since the typical concentration range of the pollutants of interest is in the order of few hundred of ppmv, direct interactions between the electrons and pollutants can usually be ignored. These NTP mechanisms are in contrast to the mechanisms involved in thermal incineration processes (plasma torches, furnaces, several chemical techniques), which require heating the entire gas stream in order to destroy the pollutants.

Pulsed corona and dielectric barrier discharge (DBD) techniques are two of the more commonly used methods for producing electrical discharge plasmas. The critical parameter in the use of these plasmas in pollution control devices is the energy cost of one removed toxic molecule. The energy cost mainly depends on the efficiency of the energy transfer from a power source to plasma reactor, configuration of electrodes, and efficiency of chemical reactions [14].

Although performances of pulsed discharges have been well demonstrated and established for atmospheric pollutant removal, some works have emphasized that combination of non thermal plasma with catalysis can be necessary to complete the cleaning process in terms of reaction by-products [15-20]). This NTP-catalyst hybrid system can enhance the removal efficiency and the reaction selectivity [4, 21-22]. A typical example is treatment of diesel and lean burn gasoline car engine exhaust for which the three-way catalyst present a low efficiency owing to the high oxygen concentration (typically 10%).

**Characteristic of NTP reactors**

Several different type plasma reactors were investigated for pollution control applications. Figure 1 shows the conventional NTP reactors that are widely used: (a) pulsed corona discharge, (b) surface discharge, (c) dielectric-barrier discharge (DBD), and (d) dielectric pellet packed-bed reactor. These reactors are subdivided according the type of discharge mode (pulse, DC, AC, RF, microwave), presence of a dielectric barrier or catalyst, and geometry (cylinder, plate). It is important to note that the chemical potential of each discharge mode differs enormously from one discharge to another. Detailed characteristics of NTP reactors may be found in [23, 24] and only a brief description of DBD reactor that we used will be given here.

The DBD reactor was originally proposed by Siemens in 1857 for "ozonizing air" [25]. The electrode configuration reactor is characterized by the presence of at least one dielectric barrier in the current path in addition to the gas gap used for discharge initiation. At atmospheric pressure, in most gases the discharge is maintained by a large number of short-lived localized current filaments called micro-discharges [26-28]. The NTP conditions in these micro-discharges can be influenced and optimized for different applications. Besides this multi-filaments mode with a seemingly random distribution of micro-discharges, regularly patterned discharges and apparently homogeneous glow discharges have been obtained in such electrode configurations. Since DBDs, can operate at high power levels and can treat large atmospheric-pressure gas flows with a negligible pressure drop, potential applications in pollution control have been investigated. Typical target substances are VOCs, such as hydrocarbons, chlorocarbons and chlorofluorocarbons, and other hazardous air pollutants. Odour control in animal houses and fish factories ($H_2S$ and $NH_3$) is another application of DBDs. Reduction of NOx in automotive diesel exhaust gases is also under intensive investigation.

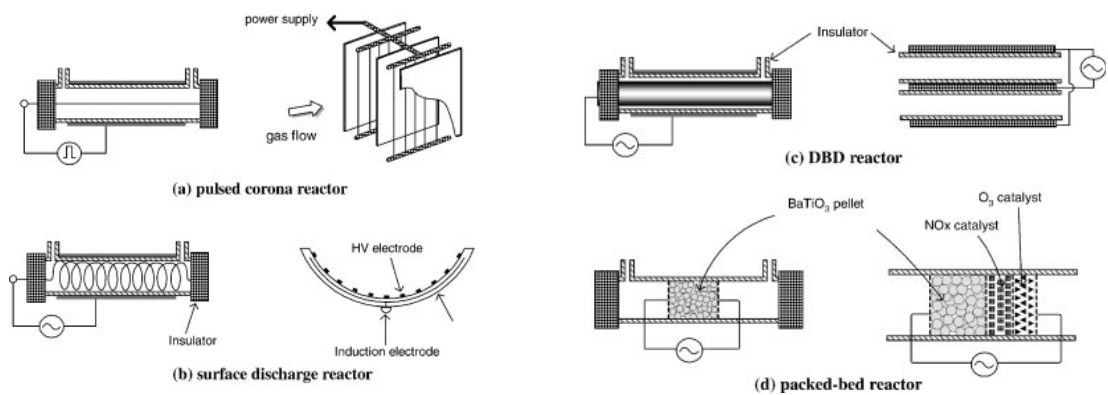

**Fig. 1.** Common DBD configurations [23]

**Removal of pollutants by NTP**

In the following we present experimental results obtained in GREMI laboratory. We used a DBD reactor in cylindrical configuration driven by a pulsed HV generator running at high repetition rate. Details of the experimental set-up are given in [8].

In the NTP, oxidation is the dominant process for exhausts containing dilute concentrations of pollutant (NO or VOC) in mixtures of $N_2$, $O_2$, with or without $H_2O$ addition. As expected, in addition of the main products of the plasma ($NO_2$, CO, and $CO_2$), a large variety of by-products are produced such as nitrite and nitrate compounds of the type R-NOx ($CH_3ONO_2$ and nitromethane ($CH_3NO_2$)). Alcohols and aldehydes are also produced following the partial oxidation of the hydrocarbon. It is emphasised in figure 2 which present infra-red measurement (FTIR) results obtained at the exit of a pulse DBD reactor for $O_2$-$C_3H_6$-NO-$N_2$ mixture.

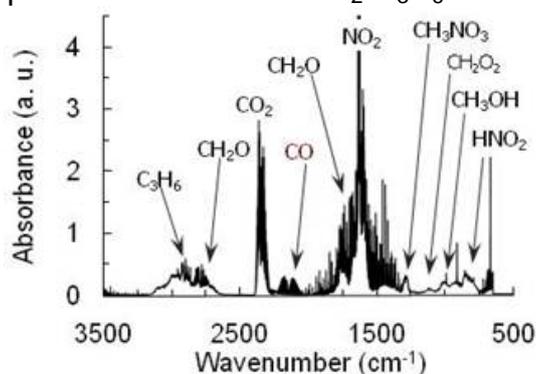

**Fig. 2.** Typical FTIR spectrum of $O_2$ (10%)-NO (500 ppm)-$C_3H_6$ (500 ppm)-$N_2$ plasma. Specific energy deposition: 27 J/L.

The process involved in NOx removal can be illustrated by examining the changes in the concentration of the NOx components (NO and $NO_2$) as the plasma specific energy is changed as shown in figure 3. In humid gas mixture, the formation of OH radicals via electron-impact dissociation of $H_2O$ and reaction of $H_2O$ with metastable oxygen atom become important and results in the formation of $HNO_2$ and $HNO_3$ by reactions with NO and $NO_2$. The role of $H_2O$ in the formation of acids becomes more apparent when $H_2O$ concentration is 5% and above. When hydrocarbons (UHC) are present in the gas mixture, OH radical becomes the main radical consuming UHC. The radical responsible for the oxidation of NO to $NO_2$ is no longer the O radical. It has been shown from detailed chemical kinetics analysis [29-30] that $HO_2$ produced from reactions involving oxidizing hydrocarbon intermediates, are the radicals that oxidize NO to $NO_2$. The OH radical reacts preferentially with the hydrocarbon (UHC), then the oxidation of NO and $NO_2$ to nitric and nitrous acids is minimized.

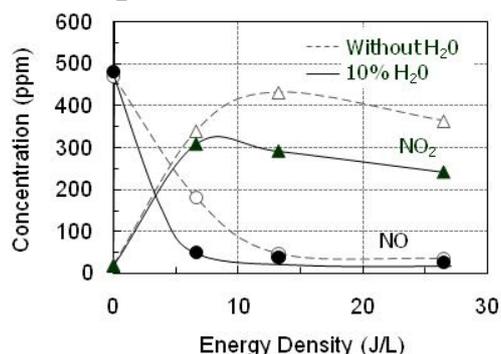

**Fig. 3.** Effect of adding 10% water vapour on concentrations NO (circles) and $NO_2$ (triangles). Gas mixture: $O_2$ (10%)-NO (500 ppm)-$C_3H_6$ (500 ppm)-$N_2$, temperature 260°C.

Experiments with gas mixtures simulating diesel (12% $O_2$, 10% $CO_2$, 10% $H_2O$, 500 ppm CO, 10% $CO_2$, 300 ppm NO, 300 ppm $C_3H_6$, $N_2$) and lean burn gasoline (6% $O_2$, 10% $CO_2$, 10% $H_2O$, 3000 ppm CO, 1000 ppm NO, 1000 ppm $C_3H_6$, $N_2$) engine exhaust were conducted as functions of temperature (150°C and 300°C). Figure 4 summarize data obtained for simulated diesel exhaust. The fraction of NOx removed significantly increases with increasing input energy density. Approximately 50% of the NOx was removed by the DBD at energy density of about 27 J/L. The fraction of NOx removed is strongly correlated to the energy cost per removed molecule. It appears that the NTP processing of diesel exhaust is much more efficient at low temperature. At 150°C, around 60% of NOx are removed at an energy cost of only 24 eV/molecule.

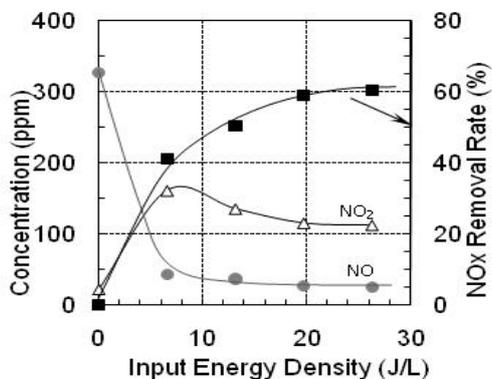

**Fig. 4.** NO, $NO_2$ concentrations and NOx removal rate as a function of input energy density in plasma processing of simulated diesel exhaust at 150°C.

**NTP-catalyst association**

The research we performed in the frame of French ECODEV program in collaboration with PSA Peugeot-Citroën and Renault companies, have demonstrated that the plasma of a sub-microsecond pulsed DBD, leading to the formation of both $NO_2$ and $C_xH_yO_z$ intermediate species at low temperature, could substitute for the first two functions of the three functions catalyst [16-17, 31]. The evaluations were made either in the absence or in the presence of plasma at 36 J/L over $Rh^{x+}/CeZrO_2$ simplified catalyst (major third function alone). As a consequence, deNOx conversion at about 270°C over simplified catalyst in the presence of plasma was found to be equivalent to three functions catalyst in the absence of plasma (figure 5). Nevertheless, the behaviour between room temperature and 170°C is quite different, due to the probable R-NOx formation in the plasma and their subsequent adsorption on the catalyst support, as evidenced by Beutel et al [32]. Figure 5 suggests an advantageous plasma-catalyst coupling effect on the NOx remediation which is in full accordance with the proposed model: deNOx conversion was about 8% without plasma and 34% NOx remediation were measured when the plasma was "on".

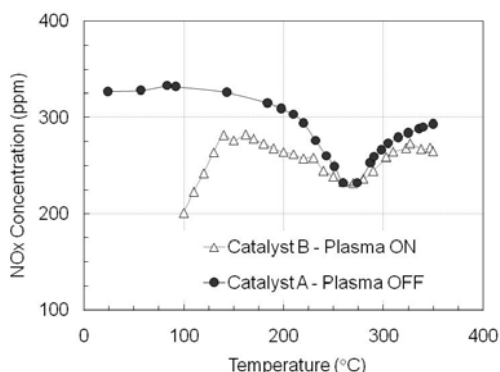

**Fig. 5.** NOx concentration at the output of plasma-catalytic reactor during a TPSR (1°C/min) of $N_2$-NO (340 ppm)-$O_2$ (8%)-$C_3H_6$ (1900 ppmC). Specific energy deposition: 36 J/L. A= Three function catalyst and B=Third function catalyst.

In the case of VOC treatment, ozone is very efficient to eliminate VOC as it has been shown for catalytic combustion of hydrocarbons and chlorocarbons. It has been also stated that the performance of NTP for the removal of VOC can be improved by the introduction of catalytic materials into the discharge zone at low temperature. A high synergy effect between catalysts and non-thermal plasma has been evidenced for some VOCs leading to a complete elimination of the pollutant molecule without appearance of undesirable by-products instead of CO and $CO_2$ [33].

A new collaboration started (mid 2008) with Professor T. Le Van (laboratory of catalysis, IAMS, Ho Chi Minh City) in order to use a pulsed DBD plasma reactor coupled to catalyst (based on noble and oxide metals) for toluene and n-hexane treatment at low temperature. The study could be extended to other molecules.

**Conclusion**

The hybrid system, NTP assisted catalytic processes, can provide a suitable alternative to reduce efficiently emissions of NOx from diesel and learn burn engine exhaust characterized by high oxygen and low unburned hydrocarbon concentrations. For some VOCs abatement studies, a high synergy effect between catalysts and NTP has been evidenced leading to a complete elimination of the pollutant molecules without product other hazardous by-products instead of CO and $CO_2$. However, Investigation of the mechanisms of catalytic reactions under plasma activation is required to improve and to optimize the hybrid system.

*Acknowledgement:*

The authors thank G. Djéga-Mariadassou and co-workers (Laboratoire de réactivité de surface, Paris VI), S. Calvo (Renault), Y. Lendresse and S. Schneider (PSA Peugeot-Citroën) for their helpful discussions.